\documentclass[a4paper,11pt]{article}
\usepackage{pos}

\title{Resolving the spacetime structure of jets with medium}
\ShortTitle{Resolving the spacetime structure of jets with medium}

\author*[a]{Adam Takacs}
\author[a]{Daniel Pablos}
\author[a]{Konrad Tywoniuk}

\affiliation[a]{Department of Physics and Technology,\\ University of Bergen, 5007 Bergen, Norway}

\emailAdd{adam.takacs@uib.no}

\abstract{
Away from the strictly soft and collinear limit of QCD radiation the choice of evolution scale in a parton shower algorithm is ambiguous and several options have been implemented in existing Monte Carlo event generators for proton-proton collisions. However, the resulting space-time evolution could result in subtle differences depending on the particular choice. In this work we quantify measurable consequences of the choice of the evolution variable and show how the implications of such a choice propagates into jet quenching observables. We develop a parton shower algorithm for a general evolution variable, that includes as special cases the virtuality, angle, transverse momentum and formation time. We study the interplay between the shower history for different evolution variables and the phase space affected by parton energy loss. In particular, we implement effects of jet quenching in the dense limit and highlight the role of color coherence effects. We compare the results of the different ordering variables to existing Monte Carlo shower implementations on the parton level by analyzing primary Lund planes. Finally, we study the sensitivity of quenched jets to the choice of evolution variable by confronting our results for a certain key observable, such as the jet mass.
}

\FullConference{%
  HardProbes2020\\
  1-6 June 2020\\
  Austin, Texas}

\begin{document}
\maketitle

\section{Introduction}
From the hard scattering partons at high energy scales are created, resulting in a big phase space for radiation due to the infrared structure of QCD. During the subsequent emissions the scale decreases until some low, nonperturbative scale is reached. This radiation process is modelled by parton showers, the essential ingredients of event generators and jet studies~\cite{Larkoski:2017jix}. 
In the last few years a lot of work was put into increasing the precision of parton showers. An important aspect of parton showers is their choice of the evolution variable, the way of treating the energy scale. 
Several choices have been made in current implementations and it is still an open question whether all of these recover the proper limits~\cite{Dasgupta:2020fwr}. It is therefore important to characterize the uncertainties emerging from the different choices. Another important choice is the momentum scheme of the splitting resulting further uncertainties to study.
Our interest in characterizing these uncertainties emerged from the medium perspective. Jets with different orderings have different spacetime structure and thus interact with the medium differently. These differences could give a hint to the spacetime structure of jets and the medium.

\section{Parton shower with different orderings}
\begin{figure}
    \centering
    \includegraphics[width=0.49\textwidth]{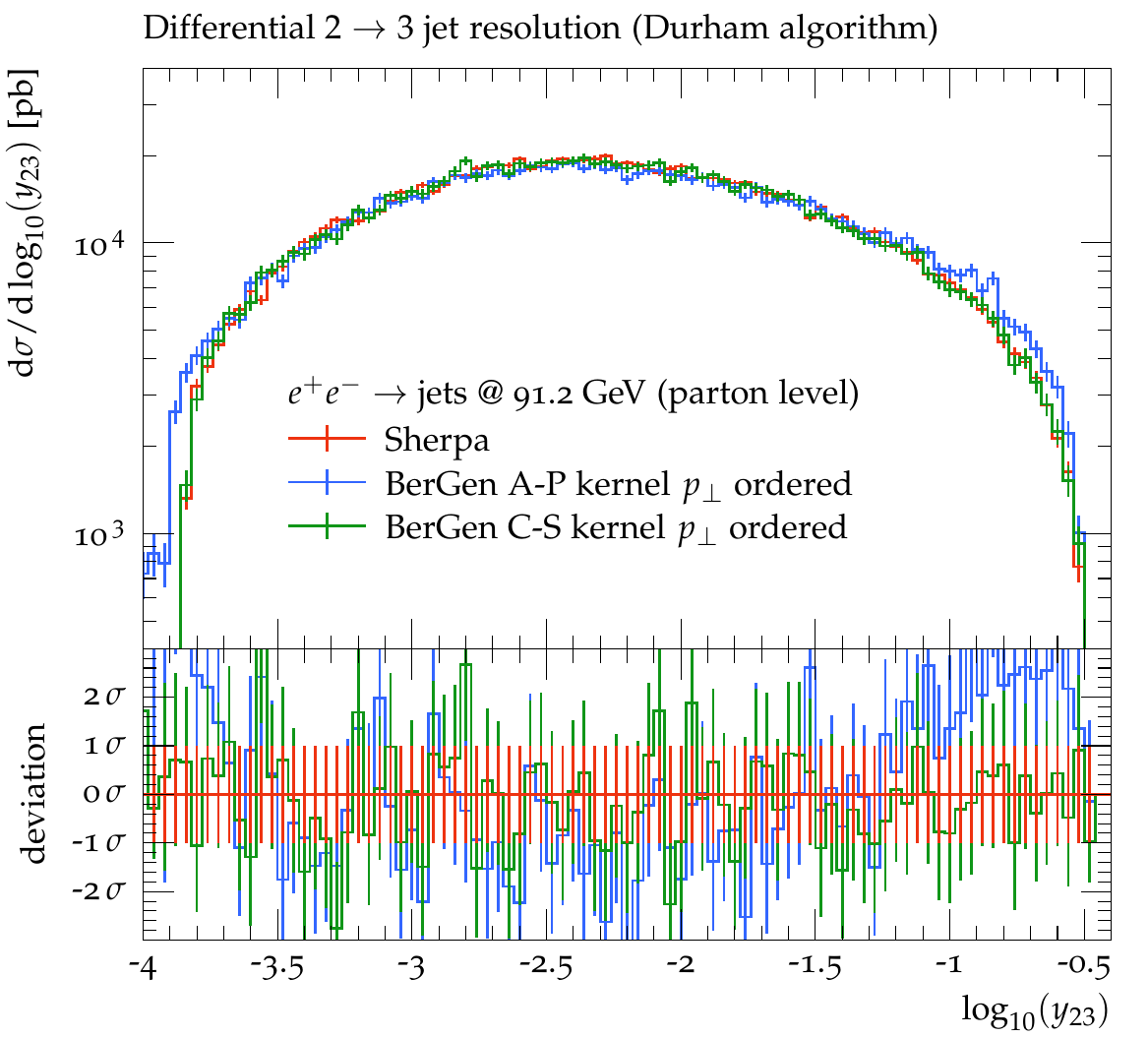}
    \includegraphics[width=0.49\textwidth]{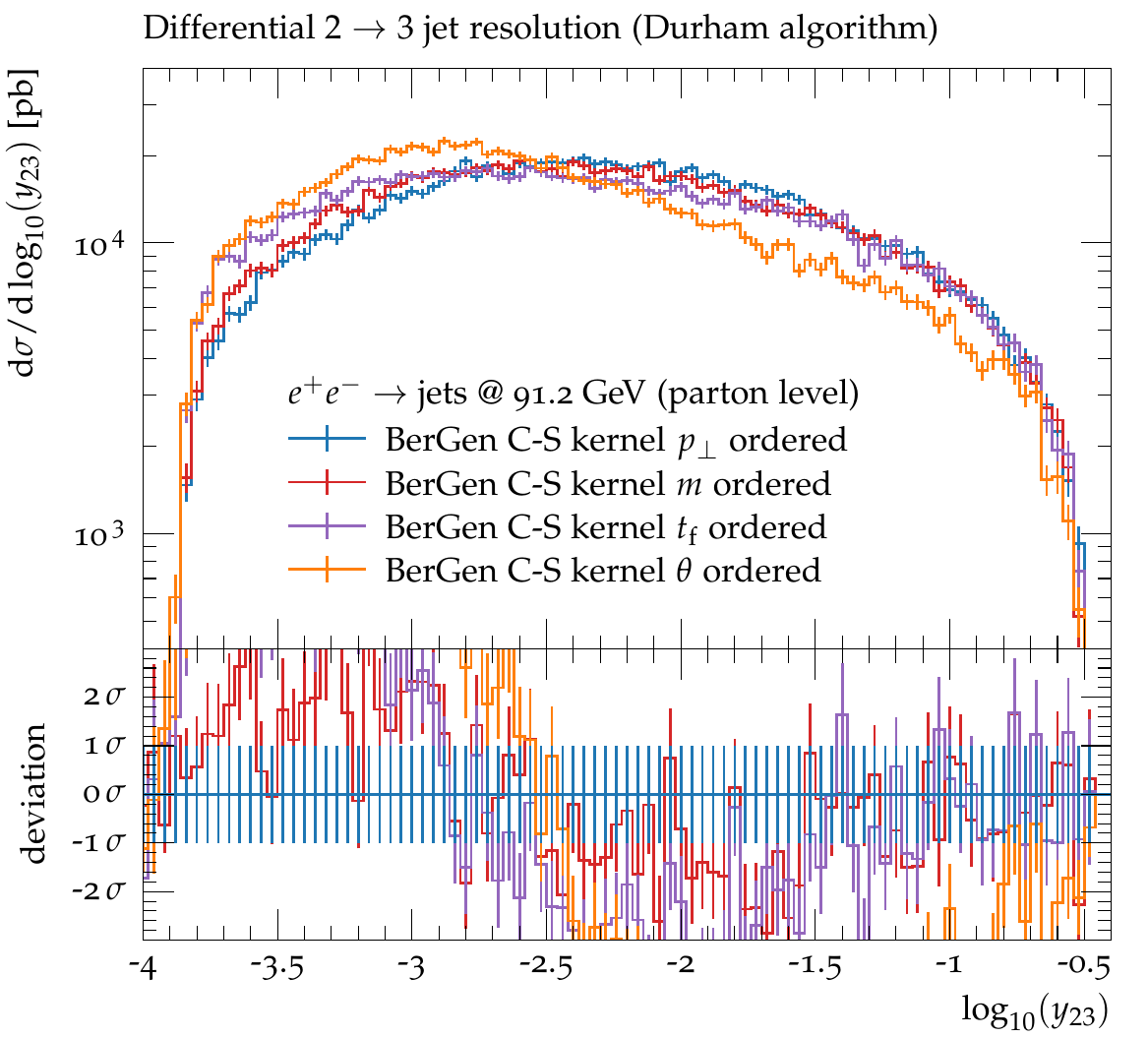}
    \caption{\textit{Left}: comparing the recoil like (blue) and the dipole (green) differential jet rate to Sherpa. \textit{Right}: the differential jet rates for different orderings in the dipole scheme.}
    \label{fig:Validation}
\end{figure}

We define a framework to implement a general ordering variable $t(Q,z,E,\dots)$. The no emission probability is
\begin{equation}
    \Delta_{\rm MC}(t)=\exp\left[-\int\frac{{\rm d}t'}{t'}\int{\rm d}z\frac{\alpha_s(p_\perp)}{2\pi}P(z,t)\Theta_{\rm MC}\Theta(t'-t)\right],
\end{equation}
where $t$ can be the mass $m$, relative transverse momentum $p_\perp$, formation time $t_{\rm f}$, energy-angle $E\theta$ or any other choice. Different momentum schemes were also implemented: PYTHIA6 like with off shell partons~\cite{Bengtsson:1986et}, PYTHIA8 recoil like with exchanged energy and on shell partons~\cite{Sjostrand:2004ef} and Catani-Seymour dipole scheme with exchanged four-momentum and on shell partons~\cite{Catani:1996vz}. All showers were initiated at the same $p_{\perp,\rm hard}$ scale and evolved down to $p_{\perp,\min}$, although the kinematic constraints built in $\Theta_{\rm MC}$ were specific for each momentum scheme. The $P(z,t)$ are the corresponding splitting functions (Altarelli-Parisi (AP) or Catani-Seymour (CS)) and the running coupling is unique $\alpha_s(p_\perp)$. We used LO $e^-e^+\to q\bar q$ matrix elements and we validated the implementations by comparing to the Sherpa event generator\footnote{We recommend S.~H\"oche note from \url{https://bitbucket.org/heprivet/mcnet-tutorial-material}}, as shown in Fig.~\ref{fig:Validation}. The closest description is given by the $p_\perp$ ordered in the dipole scheme\footnote{This is the closest to the Sherpa implementation.}, while the rest visibly deviate, especially for angular ordering.

The momentum radiated out of the jet cone is described by the quenching factor $\mathcal{Q}$ of each parton~\cite{Baier:2001yt}, which for $n$ particles modifies the spectrum as
\begin{equation}
     \frac{{\rm d}^n\sigma_{\rm med}(p_{\rm jet})}{{\rm d} p_{1}\cdot...\cdot{\rm d} p_{n}}=\mathcal Q^n(p_{\rm jet})\,\frac{{\rm d}^n\sigma_{\rm vac}(p_{\rm jet})}{{\rm d} p_1\cdot...\cdot{\rm d} p_n}\,,
\end{equation}
assuming that all branches are affected by energy loss. However, a branch is quenched only if the medium resolves its color, the splitting therefore has to be formed at time $t_{\rm f}$, before it can decohere at $t_{\rm d}$, while still inside the medium of length $L$, so $t_{\rm f}<t_{\rm d}<L$~\cite{Mehtar-Tani:2017web}.

\subsection{The Lund plane}
\begin{figure}
    \centering
    \includegraphics[width=0.32\textwidth]{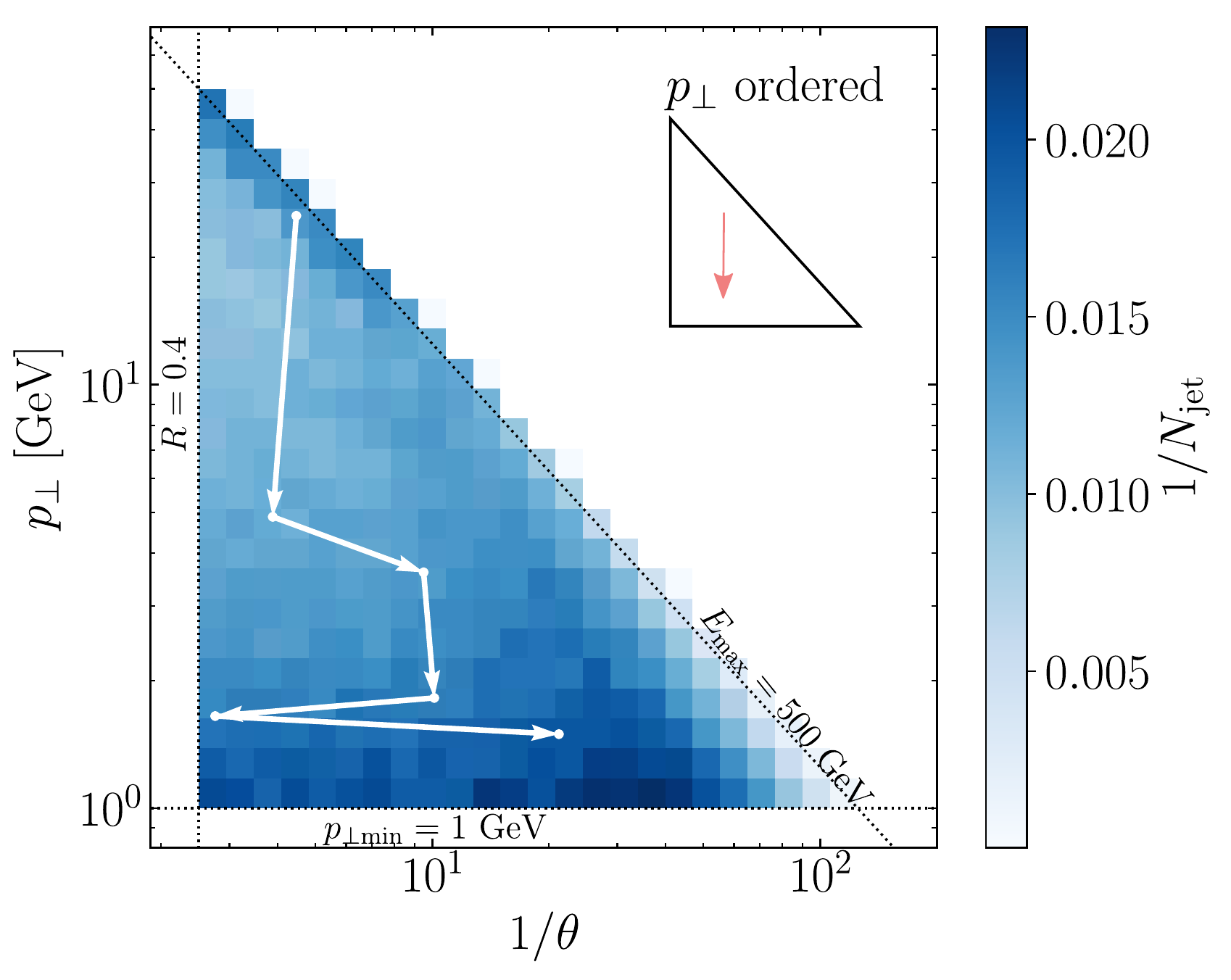}
    \includegraphics[width=0.32\textwidth]{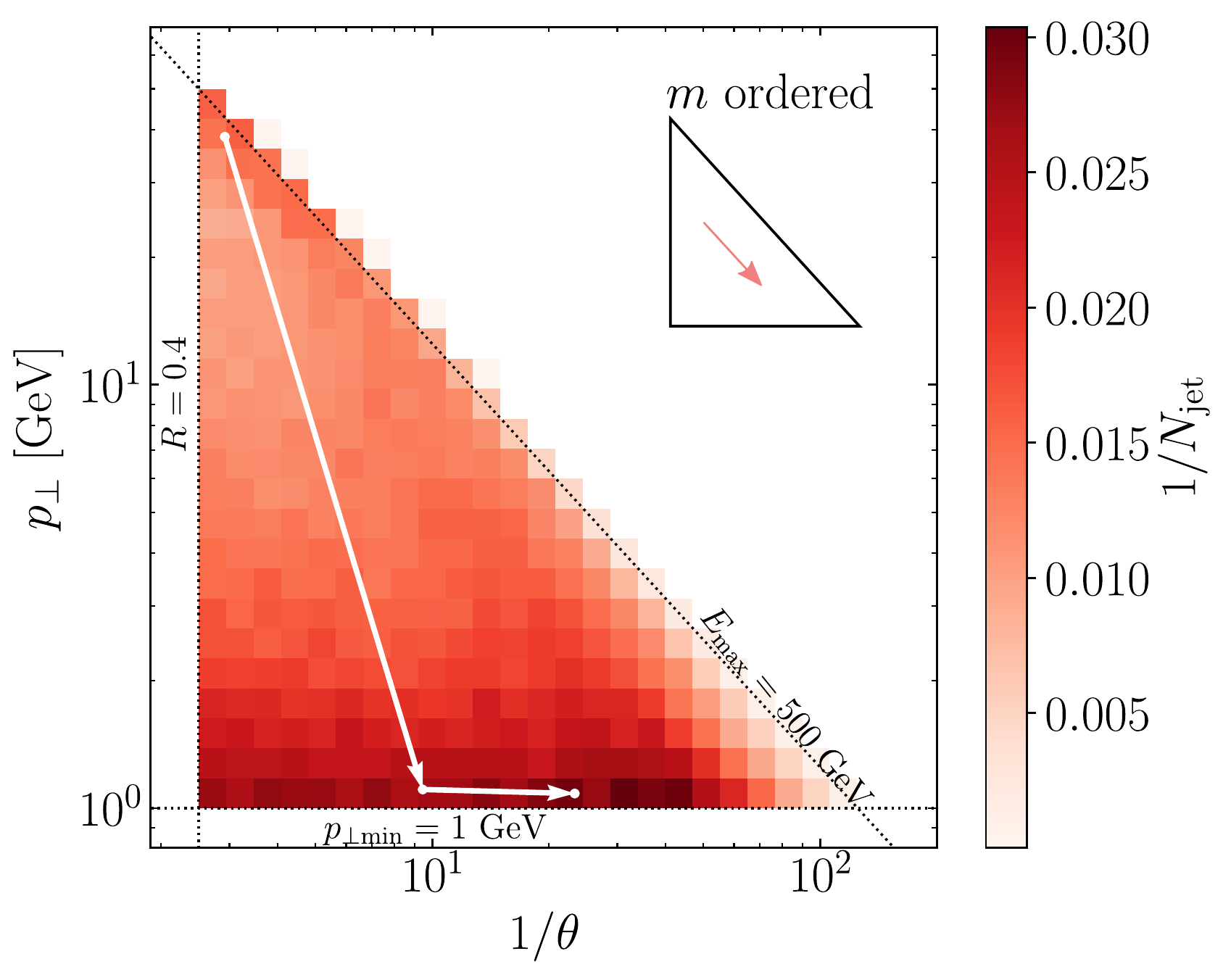}
    \includegraphics[width=0.32\textwidth]{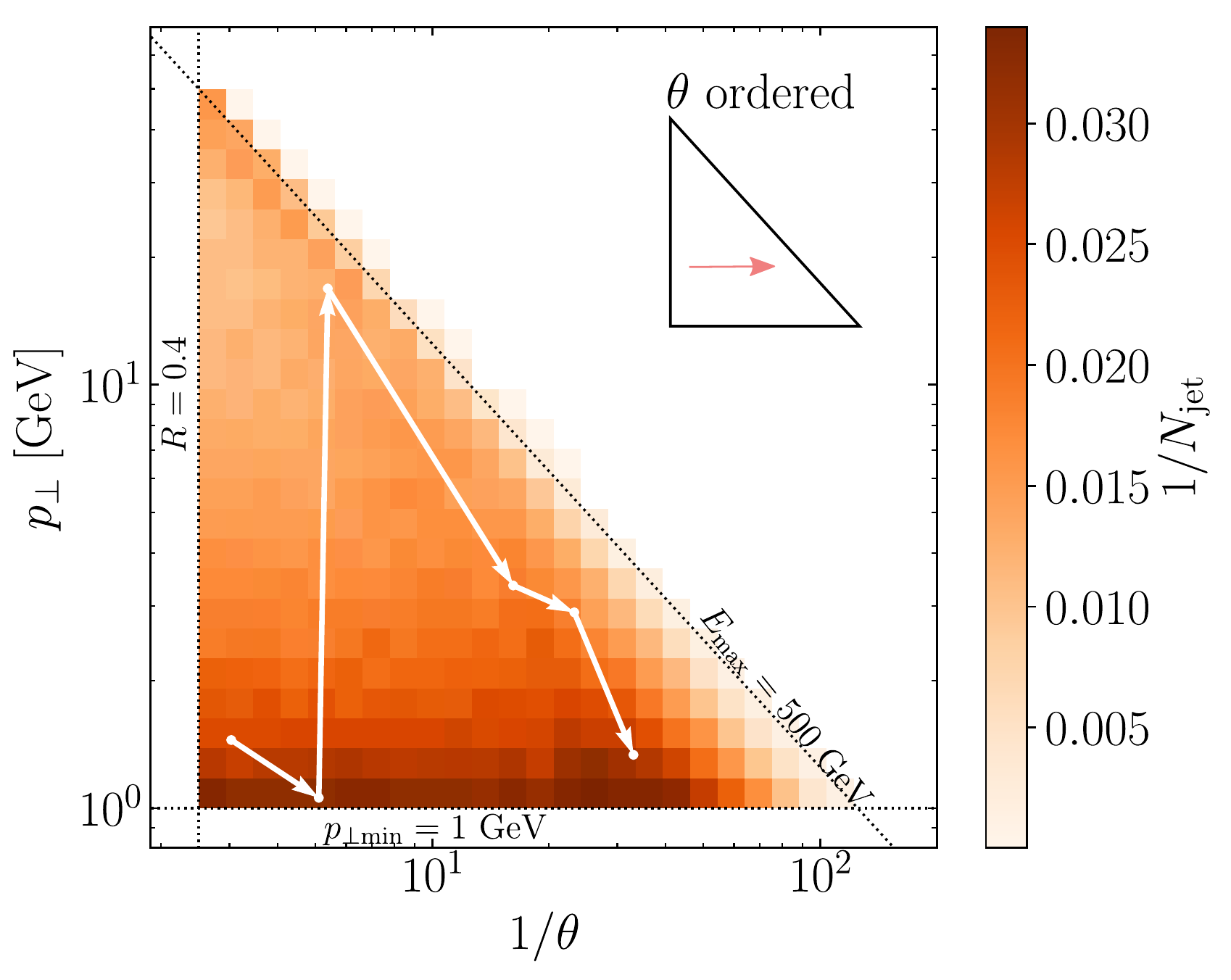}
    \includegraphics[width=0.32\textwidth]{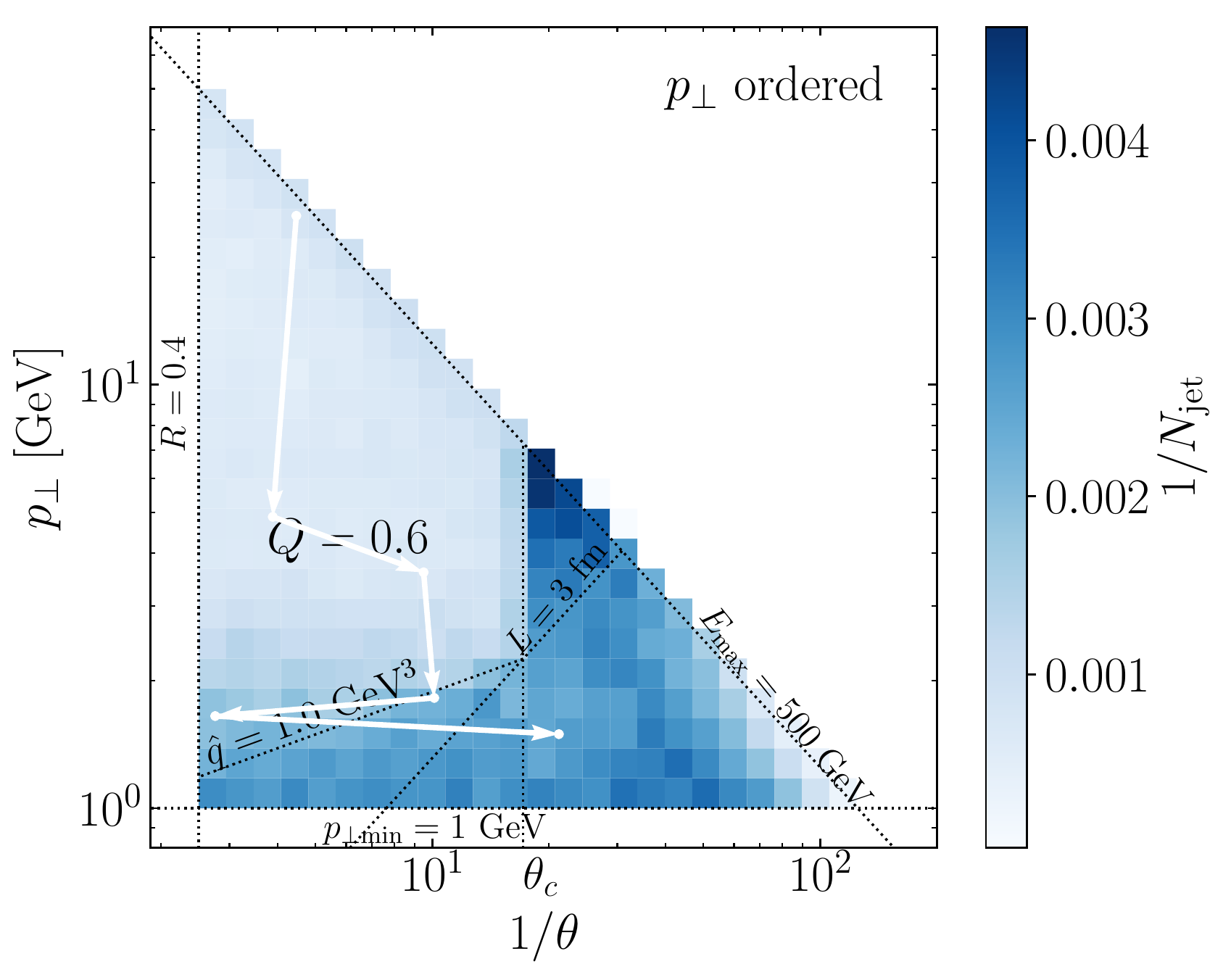}
    \includegraphics[width=0.32\textwidth]{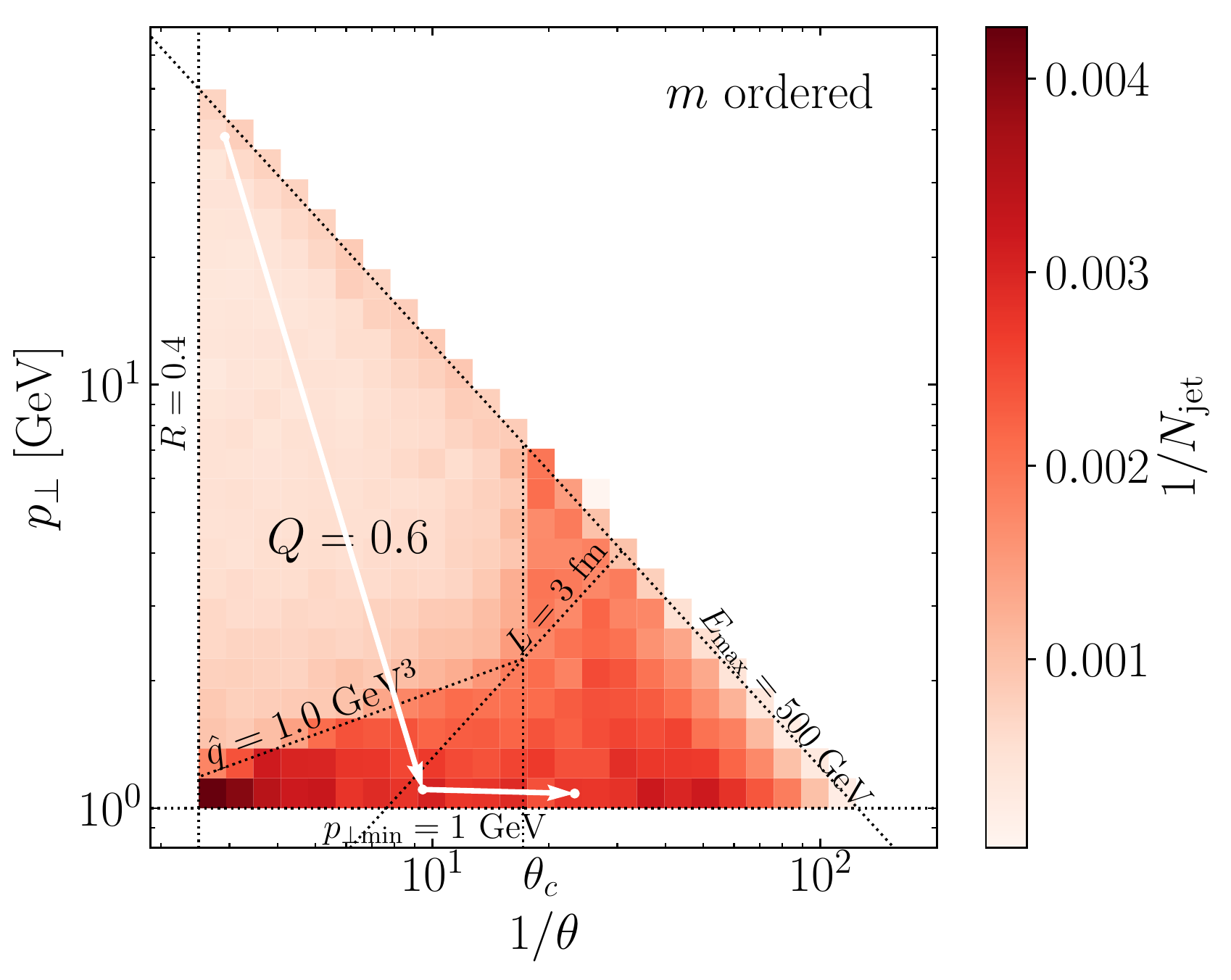}
    \includegraphics[width=0.32\textwidth]{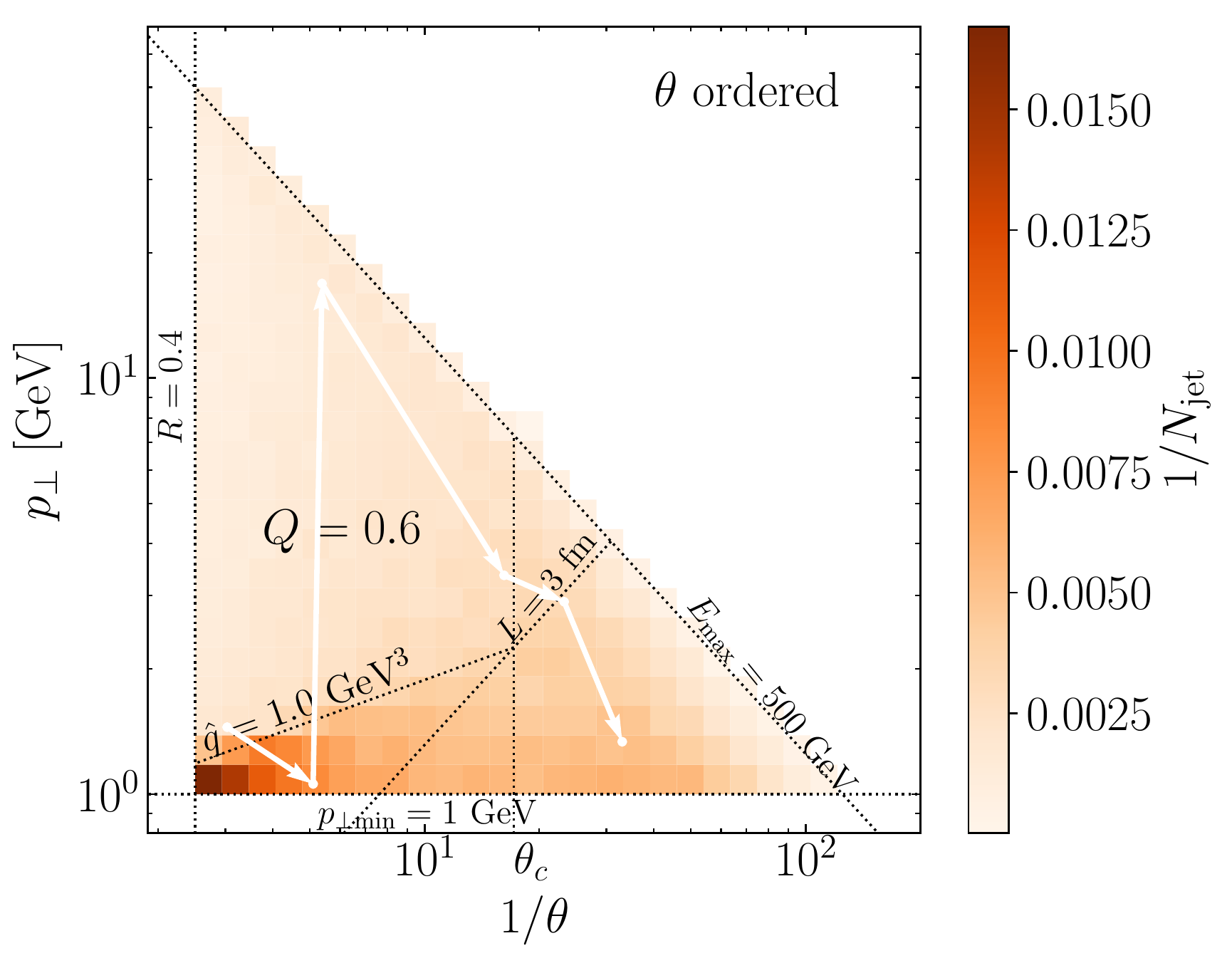}
    \includegraphics[width=0.32\textwidth]{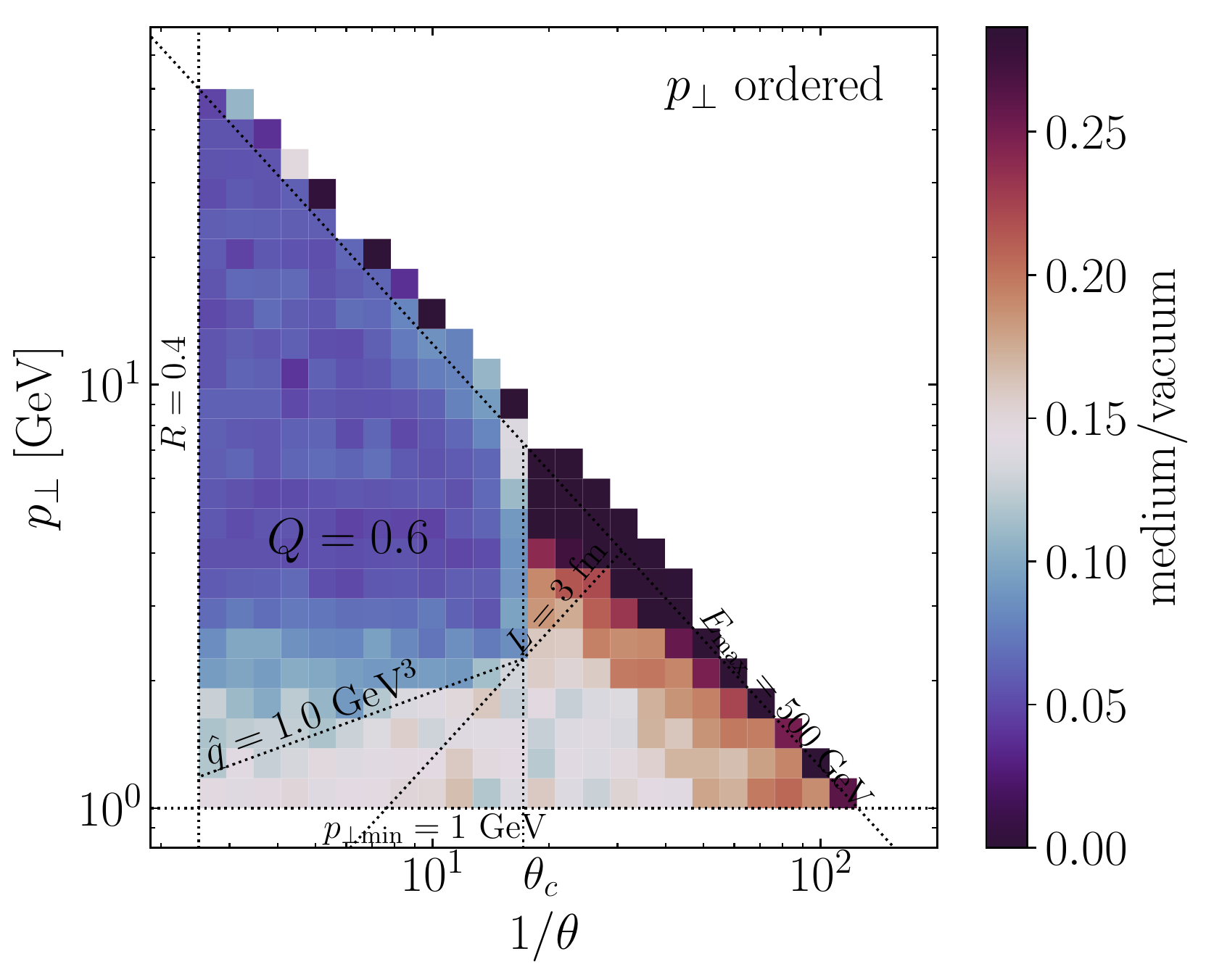}
    \includegraphics[width=0.32\textwidth]{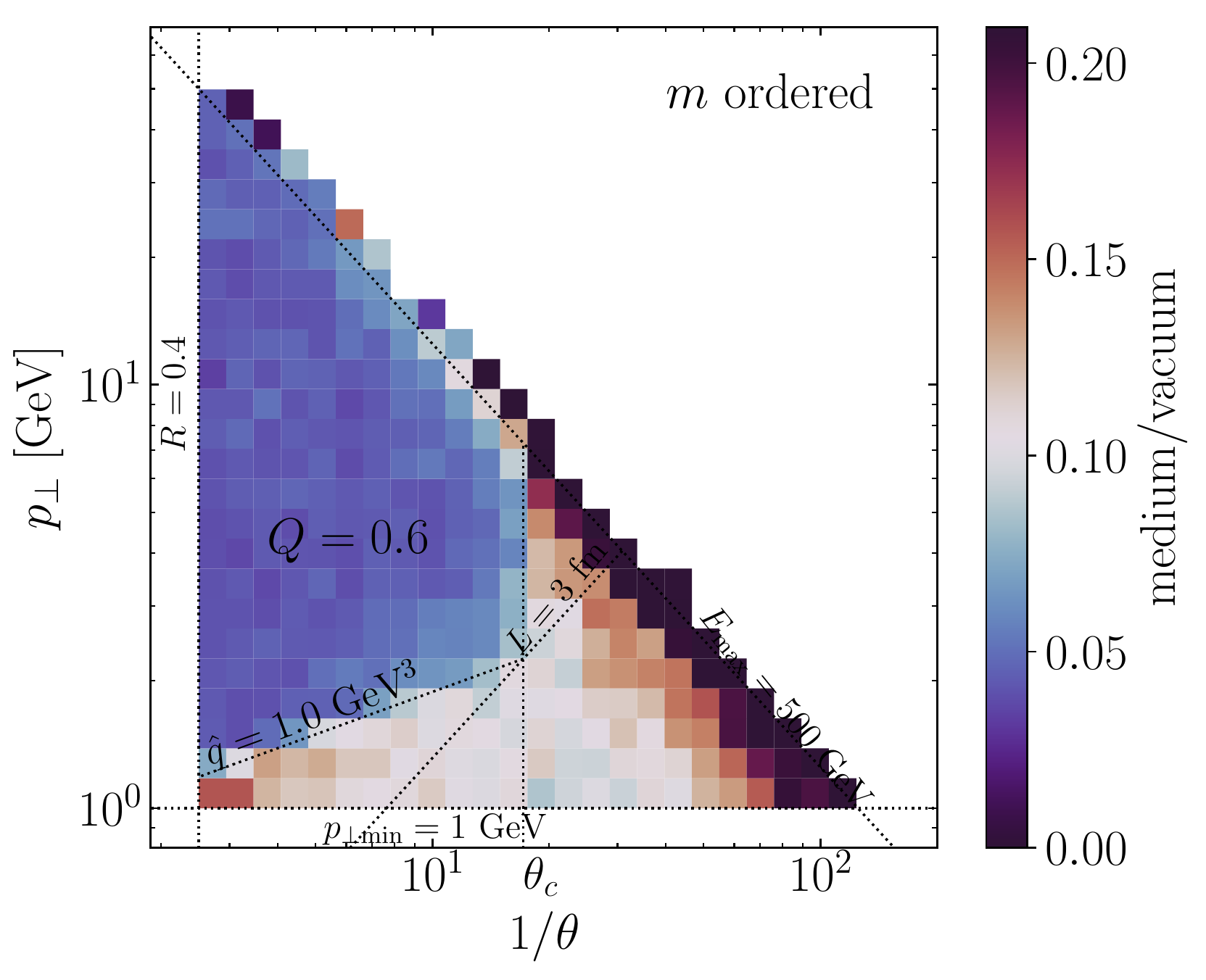}
    \includegraphics[width=0.32\textwidth]{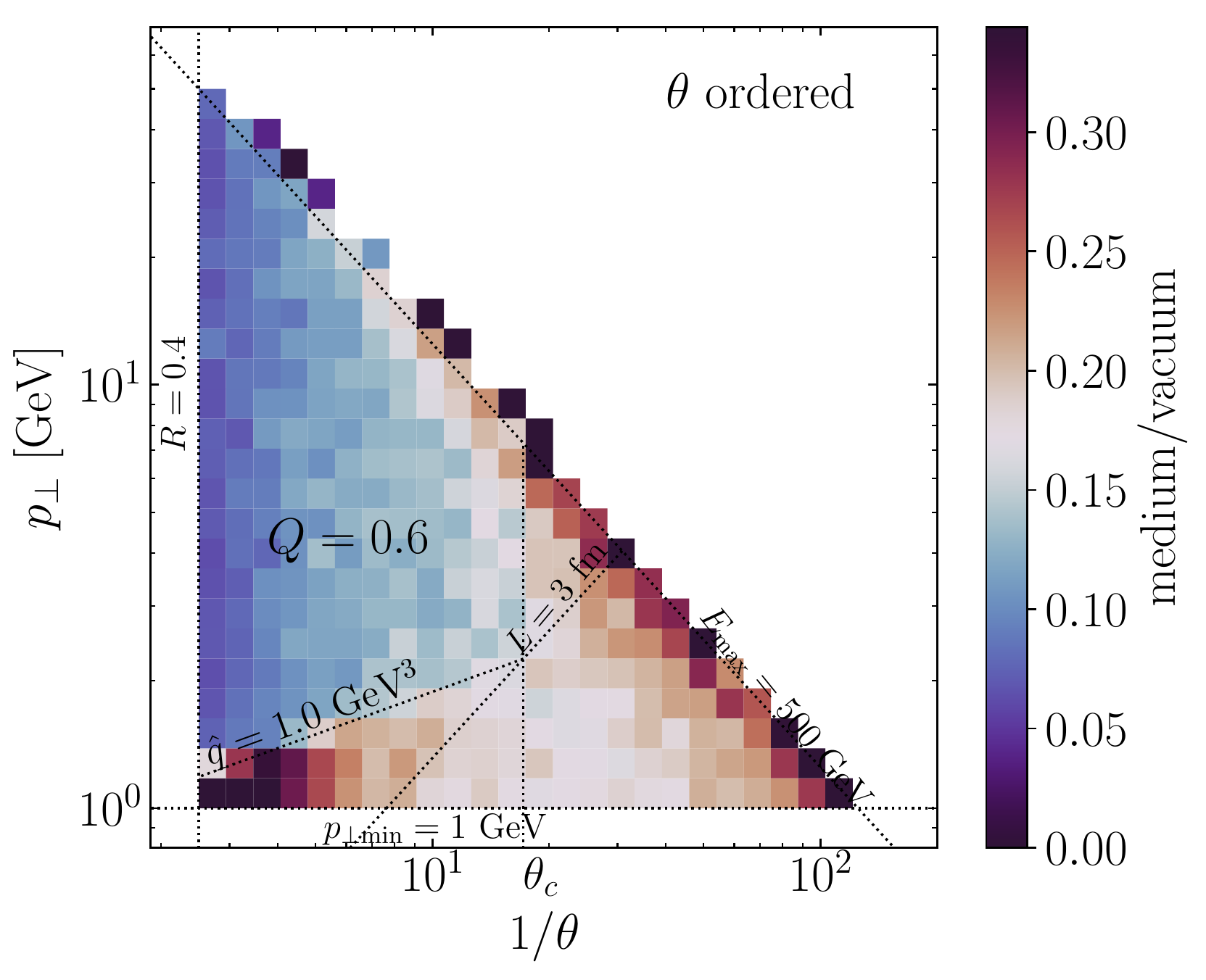}
    \caption{{\it Top}: the primary Lund planes using the PYTHIA6 like implementation. The white paths represent a jet, and the different orderings are in different columns. The pink arrows suggest the direction of evolution. {\it Middle}: the quenched results suppressing the resolved emissions. {\it Bottom}: ratio enhancing the differences.}
    \label{fig:LundPlanes}
\end{figure}

A given splitting $a\to b+c$ is described by its kinematics, for example its energy fraction $z=E_b/E_a$, the opening angle $\theta$ or the relative transverse momentum $p_\perp=z(1-z)E_a\theta$. The primary Lund plane is a 2D distribution of emissions along the hardest energy branch. It is shown for the PYTHIA6 like implementation in Fig.~\ref{fig:LundPlanes} for different orderings (upper panels). The distributions are similar driven by $\alpha_s(p_{\perp})$, but the individual jets (white paths) are ordered differently (pink arrows hint the direction). The quenched Lund planes (middle panels) show a suppression of the area populated by the resolved splittings. The ratio between the upper and middle panels highlight the differences between different orderings (lower panels). The spacetime structure is assigned to the jet using $t_{\rm f}$, and so different paths result in different ways of entering/leaving the medium.

\subsection{Jet mass}
\begin{figure}
    \centering
    \includegraphics[width=0.49\textwidth]{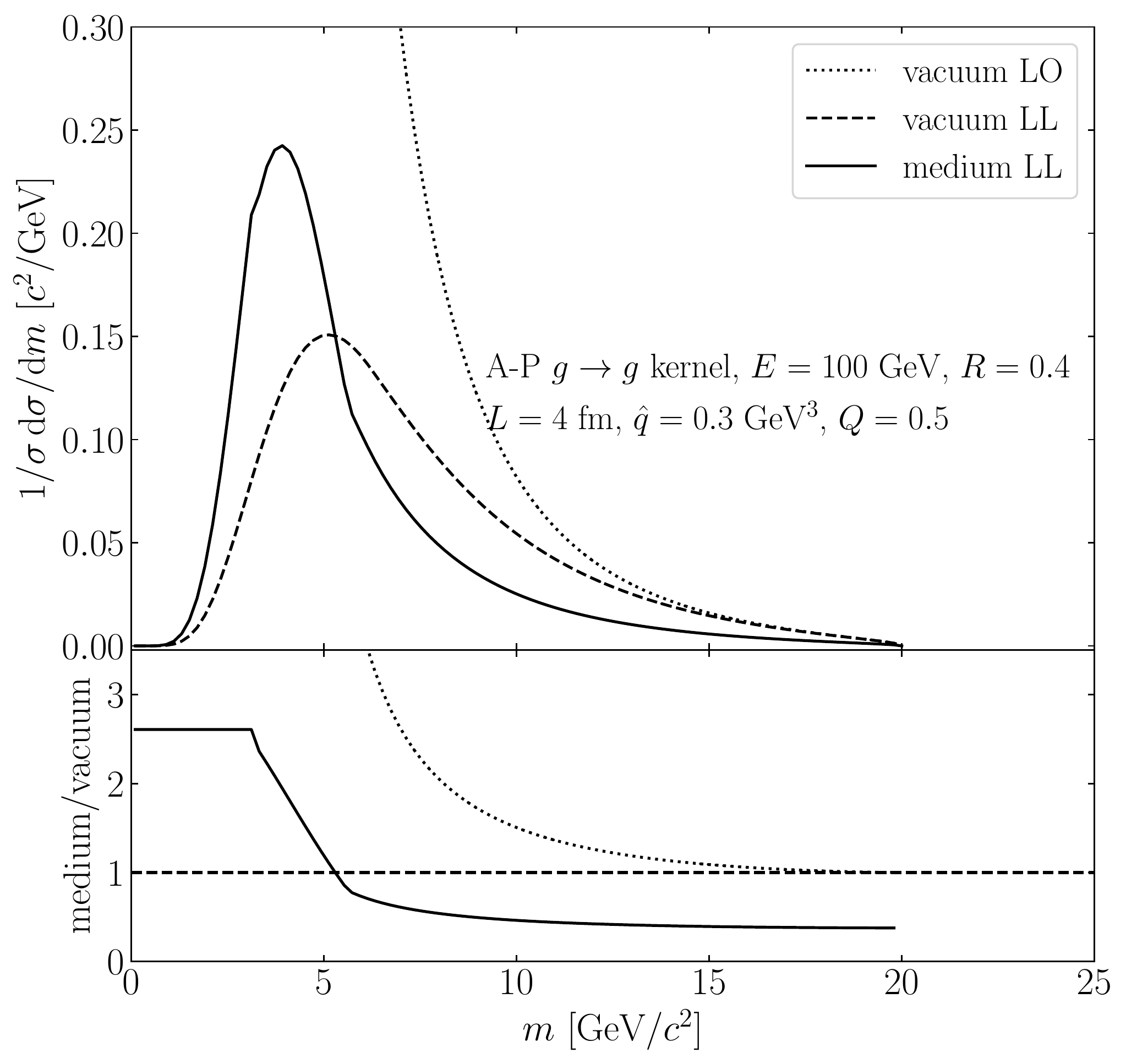}
    \caption{Jet mass in LO (dotted) and LL (dashed) accuracy for only gluons using the AP kernels for $m$ ordering. The presence of the medium is introduced through the quenching weights.}
    \label{fig:Jet_mass_LL}
\end{figure}

We choose the jet mass as an example of an observable with which to compare our different parton showers. First, we provide the LO and LL calculation only for AP gluon splittings in Fig.~\ref{fig:Jet_mass_LL}. The LL calculation for $m$ ordering shows why it is important to resum multiple emissions in order to obtain a finite peak at lower masses. We apply the medium by using the $\mathcal Q$ quenching weight
\begin{equation}
    p(m)=\int\frac{{\rm d}t}{t}\int{\rm d}z\,\Delta^{\rm med}_{\rm MC}(t)\frac{\alpha_s(p_{\perp})}{2\pi}P(z)\,\Theta^{\rm med}_{\rm MC}\,\delta\left(m(t,z)-m\right),
\end{equation}
where the splittings are quenched according to $\Theta^{\rm med}_{\rm MC}=\Theta_{\rm MC}[\mathcal Q^2\Theta_{\rm in}+\Theta_{\rm out}]$, and in(out) corresponds to the (un)resolved phase space. The result is a suppression of high masses, with an overall narrowing of the distribution. At low masses, where the emissions are not resolved by the medium, the ratio is flat.

\begin{figure}
    \centering
    \includegraphics[width=0.49\textwidth]{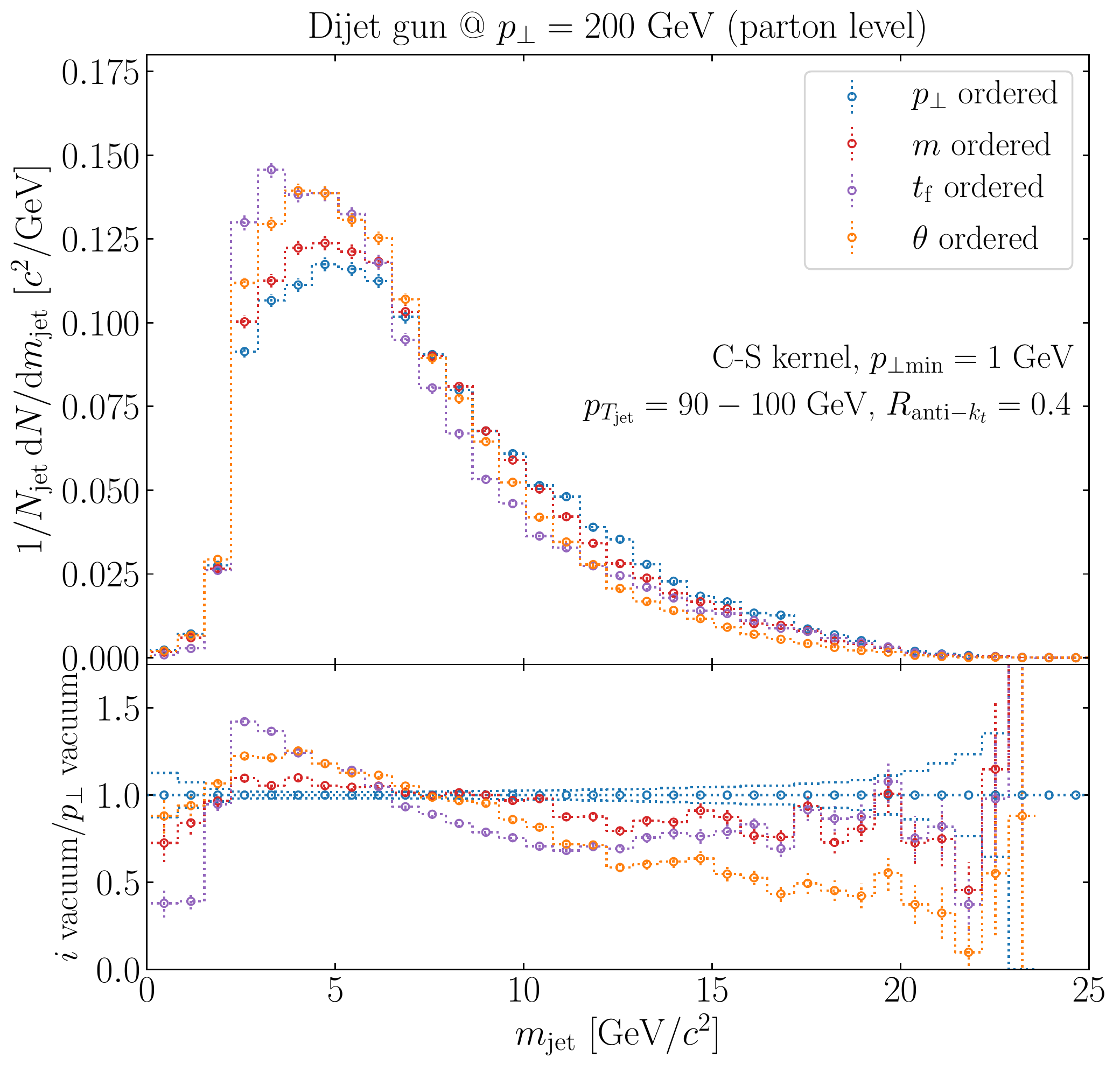}
    \includegraphics[width=0.49\textwidth]{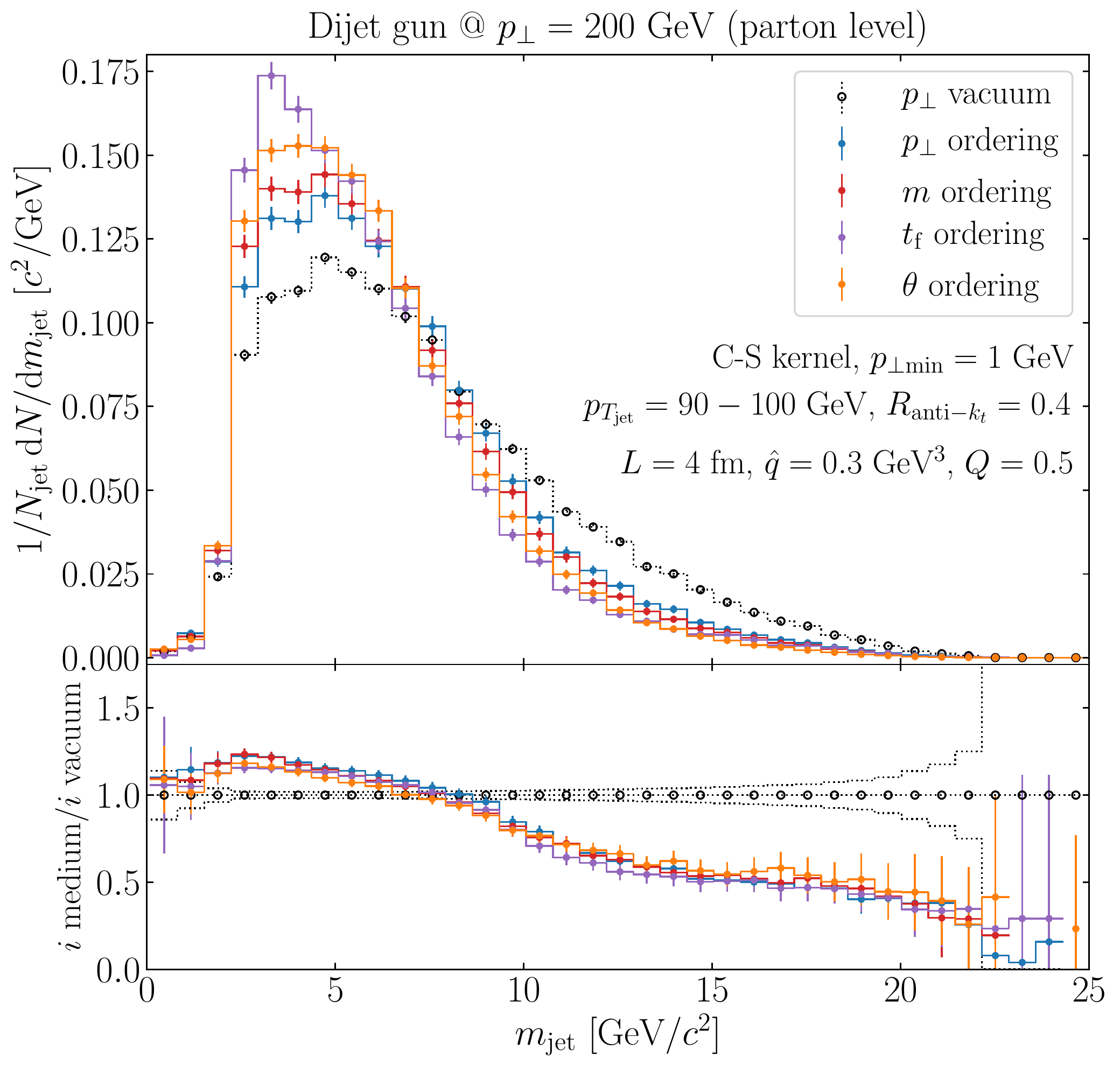}
    \caption{\textit{Left}: the reconstructed jet masses in vacuum for different orderings in the dipole scheme. \textit{Right}: the quenched jet masses. The dotted line is the $p_\perp$ ordered, vacuum result.}
    \label{fig:Jet_mass}
\end{figure}

Fig.~\ref{fig:Jet_mass} shows the reconstructed jet masses, which are similar to the LL prediction. There are non-negligible differences between the orderings in vacuum, showing the importance of characterizing these uncertainties (left panel). The right panel shows the same curves after quenching. The result is similar to the LL prediction: suppression of high masses and flattening at low masses. The individual curves differ, although the ratios (of the same ordering) fall on the top of each other.

\section{Summary}

We introduced a framework to study different orderings and momentum schemes in parton showers. We argued the importance of this for both vacuum and medium studies. Our first results on the jet mass indicate the need to understand the differences in evolution scale, even in vacuum.

\section*{Acknowledgements}
Supported by a Starting Grant from Trond Mohn Foundation (BFS2018REK01) and the University of Bergen.


\end{document}